\documentclass[a4paper,pre,twocolumn,superscriptaddress,floatfix]{revtex4-1}

\usepackage{graphicx}
\usepackage{amsmath}
\usepackage{xcolor}

\newcommand{\lvec}{\boldsymbol{l}}
\newcommand{\svec}{\hat{\boldsymbol{s}}}

\newcommand{\fp}{\color{black}}

\begin{document}

\title{Microscopic biophysical model of self-organization in tissue due to feedback between cell- and macroscopic-scale forces}

\author{J.P. Hague}
\affiliation{School of Physical Sciences, The Open University, Milton Keynes, MK7 6AA, UK}
\email{Correspondence: Jim.Hague@open.ac.uk}

\author{P.W. Mieczkowski}
\affiliation{School of Life Health and Chemical Sciences, The Open University, Milton Keynes, MK7 6AA, UK}
\affiliation{School of Physical Sciences, The Open University, Milton Keynes, MK7 6AA, UK}

\author{C. O'Rourke}
\affiliation{UCL Centre for Nerve Engineering, London, UK}

\author{A.J. Loughlin}
\affiliation{School of Life Health and Chemical Sciences, The Open University, Milton Keynes, MK7 6AA, UK}

\author{J.B. Phillips}
\affiliation{Department of Pharmacology, UCL School of Pharmacy, London, WC1N 1AX, UK}
\affiliation{UCL Centre for Nerve Engineering, London, UK}

\begin{abstract} 
We develop a microscopic biophysical model for self-organization and reshaping of artificial tissue, that is co-driven by microscopic active forces between cells and extracellular matrix (ECM), and macroscopic forces that develop within the tissue, finding close agreement with experiment. Microscopic active forces are stimulated by $\mu$m scale interactions between cells and the ECM within which they exist, and when large numbers of cells act together these forces drive, and are affected by, macroscopic-scale self-organization and reshaping of tissues in a feedback loop. To understand this loop, there is a need to: (1) construct microscopic biophysical models that can simulate these processes for the very large number of cells found in tissues; (2) validate and calibrate those models against experimental data; and (3) understand the active feedback between cells and the extracellular matrix, and its relationship to macroscopic self-organization and reshaping of tissue. Our microscopic biophysical model consists of a contractile network representing the ECM, that interacts with a large number of cells via dipole forces, to describe macroscopic self-organization and reshaping of tissue. We solve the model using simulated annealing, finding close agreement with experiments on artificial neural tissue. We discuss calibration of model parameters. We conclude that feedback between microscopic cell-ECM dipole interactions and tissue-scale forces, is a key factor in driving macroscopic self-organization and reshaping of tissue. We discuss application of the biophysical model to simulation and rational design of artificial tissues.
\end{abstract}

\maketitle

\section{Introduction}

{\fp A key factor in the self-organization and reshaping of tissues is active feedback between microscopic forces generated by cells and macroscopic tensions in the extracellular matrix (ECM).} Cells in tissues grow within ECM, a biopolymer network with viscoelastic properties. Cells generate active forces that locally distort ECM, and also respond to tensions in the ECM. We know from experiments \cite{eastwood1998}, that even in the absence of initial tension, cells in tethered artificial tissues play an active role to remodel the ECM and generate tension. On the other hand, cells respond to this tension and align in a feedback loop. Thus, tissues are active materials that manifest self-organized order and spontaneous symmetry breaking due to the local reactions of cells to the properties of their environment and vice versa. If sufficient numbers of cells act in the same way, the resulting macroscopic tension reshapes the tissue, and also directs self-organization in the tissue, creating a microscopic-macroscopic feedback loop.  Tension fields generated by these microscopic distortions and macroscopic reshaping of the ECM, lead to regions of both self-organization and disorder within engineered tissue samples \cite{eastwood1998,obrien2011,orourke2015,orourke2017}.

{\fp The goal of this paper is to obtain greater insight into the processes of ECM-cell interaction, by developing a {\bf microscopic} biophysical model that includes those interactions on a cellular scale, yet which is sufficiently simple to describe how these microscopic processes affect, and are affected by, self-organization on the macroscopic length scales of tissue.} Therefore, we aim to develop a biophysical model that (1) is sufficiently complex to incorporate the physics of the subtle interplay between the extracellular matrix and active forces in cells; (2) can simulate the reshaping process and self-organization on a macroscopic scale; and (3) is sufficiently simple to describe large numbers of cells in tissue structures. The alignment of cells in tissues is similar to the nematic order found in liquid crystals, raising the possibility that microscopic models of the type used in statistical physics could be used to simulate artificial tissues. Our approach is different to previous work, with emphasis on a mixed phenomenological and biophysical model that is microscopic, yet fast to solve so that predictions can be made for the behavior of extremely large numbers of cells in tissues. 

{\fp Continuum models have been developed to describe the alignment of stress fibers in artificial tissue constructs through both passive elasticity and the active forces generated by cells \cite{deshpande2006,pathak2008,obbinkhuizer2014}.} Legand {\it et al.} \cite{legand2009} have compared the continuum model of Refs. \cite{deshpande2006,pathak2008} with artificial tissue structures grown in MEMS systems, finding good agreement between model and experiment.  Continuum models necessarily coarse grain cells and ECM into a single effective material containing the properties of both cells and ECM. Once this has been done, cells cannot be individually tracked, and cell alignments are assumed to follow stress fibers directly. A description of tissues as active continuum materials is well suited for modeling situations where the properties of individual cells are not important.

{\fp Microscopic models are complementary, but distinct to, continuum approaches.} In a microscopic model, it is possible to model the complex self-organization of an ensemble of distinct cells in response to their environments, without the coarse graining associated with continuum models, and thus allowing simulations to be carried out that treat processes on cellular length scales. On the other hand, a typical challenge for microscopic models is carrying out simulations that are large enough to capture macroscopic scales.

{\fp To develop a microscopic model of ECM-cell interactions leading to self-organization, that can be simulated on macroscopic scales, we take inspiration from force-dipole models and contractile network models.} A force-dipole approach could be used to simulate the interactions between cells and the ECM (see e.g. \cite{schwarz2013}). Since it has been established that cells find an orientation consistent with the minimum work needed to obtain that configuration \cite{bischofs2003}, it would then suffice to find the lowest energy (ground) state of the theoretical model. However, the cells in such force-dipole models must be constrained to occupy nearly-fixed positions in their substrates to avoid volume collapse relating to the form of the interaction (which has a $1/r^3$ attractive form). Thus, such models cannot describe the reshaping of tissue that occurs simultaneously with self-organization. We also take inspiration from contractile network models, in which networks of springs can be used to model elastic effects in e.g. biopolymers \cite{guthardttorres2012,boal}. To date, the active forces due to cell-ECM interaction have not been introduced into contractile network models.

{\fp Our COntractile Network Dipole ORentiation (CONDOR) model of cellular self-organization goes beyond previous work by combining force-dipole orientation models with contractile network models.}  Our microscopic model combines a complex network of elastic fibers that represents the ECM, with active forces that follow cell symmetries, to describe the active feedback between cells and ECM. This active feedback leads and responds to the macroscopic self-organization and reshaping of tissues. It is able to describe the tension fields, self-organization, and reshaping of tissues that arise from the active forces between cells and ECM.  Spring networks are a popular way of representing elasticity in biopolymers \cite{boal}, and force-dipoles are a successful way of representing ordering of cells in tissues \cite{schwarz2013}, but the two approaches have not yet been combined in this way.

{\fp Beyond the biophysical description of complexity in tissues, our CONDOR model could assist in the rational design of the scaffolds and molds that direct self-organization in engineered (or artificial) tissues.} Artificial tissues are important in both pharmaceutical testing and regenerative medicine. In order to be useful for these applications, engineered tissues need to mimic the properties of real tissues. For example, it is critical that cells in artificial nervous and corneal tissues are  self-organized so that they are highly aligned \cite{georgiou2013,phillips2005,mukhey2018,east2010}. 
The development of new molds and scaffolds to encourage biologically relevant self-organization can take many months or years of painstaking trial and error. Thus, there is a need for new biophysical models capable of simulating the onset of self-organization in artificial tissues, which can be used for predicting which mold and scaffold designs have a good chance of success. Hence, our secondary goal is to generate techniques for detailed large-scale simulations of self-organization in artificial tissues that can be used as tools to assist rational design of molds and scaffolds.

{\fp This paper is organised as follows. In Sec. \ref{sec:method}, the CONDOR model and simulated annealing method for solving it are introduced.} We present results from our microscopic biophysical model in Sec. \ref{sec:results}, which also includes comparisons between simulation results and existing experimental data. Finally, in Sec. \ref{sec:discussion} we conclude and discuss the context of our microscopic biophysical model, the validation and calibration of our model, how our simulations could be used for rational design of artificial tissue molds, and finally the outlook for extensions of the model.

\section{Method and CONDOR model}
\label{sec:method}

{\fp In this section, we develop the CONDOR biophysical model that we use to describe how the active feedback between cells and ECM leads to reshaping of, and self-organization in, tissues; discuss how the parameters of the CONDOR model can be related to the cell density; and describe details of the simulated annealing method that we use to solve the CONDOR model.}

\subsection{Development of the CONDOR model for self-organization in engineered tissue}

\begin{figure}
\includegraphics[width=85mm]{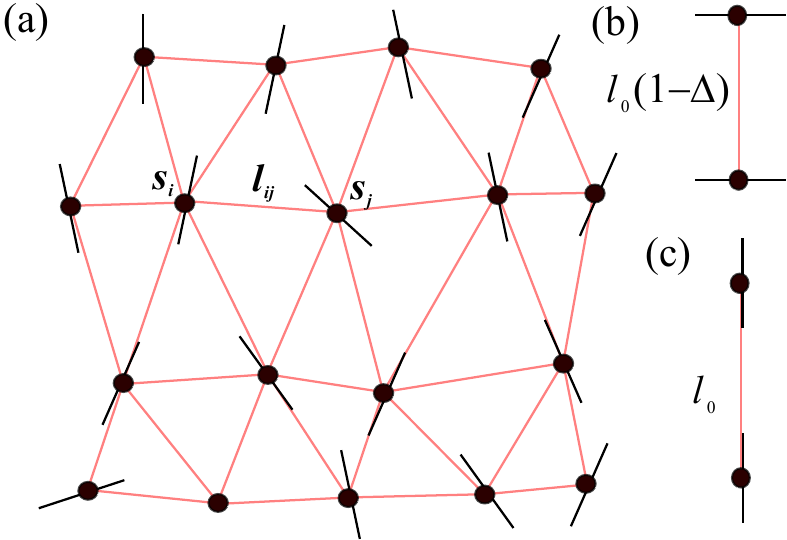}
\caption{(a) A two-dimensional schematic of the three-dimensional
  CONDOR model. Springs represent the extracellular matrix. Cells at
  vertices are indexed as $i$ and $j$. The distance between cells is
  $\lvec_{ij}$, and cell orientation at each vertex is labeled
  $\svec_{i}$; (b) If cells are perpendicular to a bond, then cell
  generated forces reduce equilibrium length by $l_{0}\Delta$; (c) If
  cells are parallel to the bond, then equilibrium length is unchanged
  at $l_0$. 3D calculations presented here use FCC networks with next
  and next-next nearest (NN and NNN) neighbor spring constants
  $\kappa_{\rm NNN}/\kappa_{\rm NN}=0.4$.}
\label{fig:model}
\end{figure}

{\fp The biophysical model developed here is based on a three-dimensional network of
springs representing the ECM; with the interactions between cells and ECM introduced by changes in the undeformed length of springs dictated by the symmetry of the cells.} The three-dimensional spring network has a face-center-cubic (FCC) geometry. The cells in the artificial tissue structures that we simulate here are well separated, such that there is no direct interaction between cells; rather, the interaction is mediated via the springs representing ECM. The glial cells considered here (and many other cell types) are naturally elongated, and thus assumed to have approximately order 2 rotational symmetry.  This is represented by an order 2 correction to the undeformed length of springs consistent with a force dipole. The network may, or may not, be subject to external stresses (depending on the system to be simulated). In all the cases considered here, the network is not subject to external stresses. A schematic of the model is shown in Fig. \ref{fig:model}.

{\fp Discussion of the form of the model begins by considering the energy of the contractile network representing the ECM in the absence of
cells, which is given by a sum over Hooke's law
for each individual filament,
%
$E = \frac{1}{2} \sum_{ij} \kappa_{ij} (|\lvec_{ij}| - l_{{\rm u},ij})^2$.}
%
Here, $\lvec_{ij}$ is the displacement between cells at vertices $i$
and $j$, $l_{{\rm u},ij}$ is the undeformed length of springs and
$\kappa_{ij}$ is the spring constant. In the absence of cells and
external tension, the network is stress free. In all the examples
considered in this paper, there is no external tension, so the network
is stress free in the absence of cells. In the cases we consider,
cells generate stress in the network. However, it is also possible to
introduce stress by pulling on tethers that hold the tissue in place.

{\fp We introduce the effect of the active forces generated between cells and the network, via a modification of the undeformed length of the neighboring springs, $l_{{\rm u},ij} = l_{0,ij} - f\left(\Theta_{1,ij},\Theta_{2,ij}\right)$.} $l_{0,ij}$ is the undeformed length of the bond in the absence of cells, $\Theta_{1,ij}$ and $\Theta_{2,ij}$ are the angles between a bond and cell orientations at the end of the bond. For simplicity, cells are positioned at the vertices between springs. Cells are able to remodel, pull on and respond to the extracellular matrix \cite{kollmannsberger2011}, and as such the tissue sample, which is a material comprising both ECM and cells, is active and can generate its own stresses. It is noted that the modification we consider here represents the end result of the active processes, rather than the time-dependent details of such processes.

{\fp A form of $f$ consistent with order 2 symmetry is,}
\begin{equation} 
f\left(\Theta_{1,ij},\Theta_{2,ij}\right) = l_{0,ij}\frac{\Delta_{ij}}{2}\left(2 -
\cos^2(\Theta_{1,ij}) - \cos^2(\Theta_{2,ij})\right), 
\end{equation} 
where $0<\Delta<1$ is a dimensionless constant controlling contraction
perpendicular to the long axis of the cell. For each bond,
$\cos(\Theta_{ij}) = \hat{\lvec}_{ij}\cdot \svec_{i}$, where $\svec_{i}$ is
a unitary vector representing the orientation of the cell and
$\hat{\lvec}=\lvec_{ij}/|\lvec_{ij}|$. There are two length corrections, since there are cells at both ends of the bond (spring).

{\fp Thus, taking into account the effects of corrections to the length due to active forces generated by cells, the total energy of the CONDOR model is,} 
\begin{widetext}
\begin{equation} 
E = \sum_{ij} \frac{\kappa_{ij}}{2}\left( |\lvec_{ij}| - l_{0,ij}\left(1-\frac{\Delta_{ij}}{2}\left(2-|\hat{\lvec}_{ij}\cdot\svec_{i}|^2-|\hat{\lvec}_{ij}\cdot\svec_{j}|^2\right)\right)\right)^2
\end{equation}
\end{widetext}
A schematic of the CONDOR model can be seen in
Fig. \ref{fig:model}.

{\fp $\Delta$ is a microscopic phenomenological parameter, which contains information about the level of contraction in the ECM caused by the activity of cells.} It represents a drawing in of the ECM perpendicular to the orientation of the cell. Thus, it relates the active forces in the cells to the physical system of springs representing the ECM that connects the cells. We are not able to derive a form for this parameter, since it depends on cell specific processes. In Section \ref{sec:deltaparameter}, we shall show how it is possible to establish a generic relation between the cell density and $\Delta$, by introducing a single free parameter that can be determined from comparison with experiment, such that predictions can be made.

{\fp The spring network used here is constructed on a regular lattice, and it is important to select a lattice with appropriate shear moduli to represent the ECM.} Spring networks constructed from square or simple cubic lattices
have zero shear modulus \cite{boal}. Thus our fully three dimensional simulations are carried out using face-centered-cubic (FCC)
lattices of cells, since the tetrahedral structures in these lattices ensure
shear modulus. The shear and bulk moduli of the ECM,
can be controlled more subtly through the ratio of next-nearest neighbor (NNN) to near
neighbor (NN) coupling, $\kappa_{\rm NNN}/\kappa_{\rm NN}$. Moreover, the inclusion of $\kappa_{\rm NNN}$ improves the stability of the spring network against volume collapse.

\subsection{Relating the CONDOR model to artificial tissue}
\label{sec:deltaparameter}

{\fp In artificial tissue experiments, where cell-laden hydrogels are cast into molds, two key parameters control how cells interact with the ECM: (1) the cell density in the hydrogel and (2) the density of biopolymers in the hydrogel (for example collagen).} In this section, we describe how parameters of the CONDOR model and artificial tissue can be related.

{\fp Samples grown in cylindrical wells undergo larger contraction as seeding density increases, until contraction saturates at high densities \cite{orourke2015}.} To understand why contraction of artificial tissue samples saturates at large seeding densities, we consider how regions within which cells influence the ECM overlap as density increases. If density is low, the volume of influence increases linearly with cell density. However, at high density the volumes of influence around each cell start to overlap, so increasing the density further does not increase the overall volume of ECM that is influenced by the cells.

{\fp The influence of this overlap on the contraction can be estimated for a random distribution of cells, and used to arrive at a functional form to relate cell density, $\rho$, to $\Delta$.} We denote the volume of influence for a single cell as $V_{\rm infl}$, and the volume of influence of all $n$ cells as $V_{{\rm tot},n}$. Note that $V_{{\rm tot},n}<nV_{\rm infl}$, since the volumes of influence can overlap when cells are densely packed. 


{\fp The mean additional volume of ECM, influenced by adding the $n$th cell, is $V_{\rm infl}(1-V_{{\rm tot},n-1}/V_{0})$, where $V_{0}$ is the total volume of the tissue.} This can be determined by considering the effect of adding an additional cell to a set of $n-1$ cells. When the $n$th cell is added, the probability that its volume of influence overlaps with that of existing cells is $V_{{\rm tot},n-1}/V_{0}$, which is the ratio of $V_{{\rm tot},n-1}$ (the volume of influence of the existing $n-1$ cells) and $V_{0}$. Thus, the probability that the volume of influence of the new cell does not overlap is $(1-V_{{\rm tot},n-1}/V_{0})$, and the mean additional volume of ECM influenced by the new cell can be calculated.

{\fp Thus, a recurrence relation is generated for the $n$th term in the expansion of the total volume of influence, $V_{{\rm tot},n}$,}
\begin{eqnarray}
  V_{{\rm tot},n} & = & V_{{\rm tot},n-1}+V_{\rm infl}\left(1-\frac{V_{{\rm tot},n-1}}{V_{0}}\right)\\
  & = &  V_{{\rm tot},n-1}\left(1-\frac{V_{\rm infl}}{V_{0}}\right)+V_{\rm infl}\\
  & = & V_{{\rm tot},n-1} R + V_{\rm infl}
\end{eqnarray}
where we define, $R=1-V_{\rm infl}/V_{0}$, $V_{{\rm tot},0}=V_{\rm infl}$. Repeated iteration then leads to a geometric series. Since $R<1$ this series can be computed analytically for the volume of influence of $N$ cells as,
\begin{equation}
  V_{\rm infl,tot} = V_{\rm infl}\left(\frac{1-R^{N}}{1-R}\right).
\end{equation}
So, assuming that $\Delta\propto V_{\rm infl,tot}$, and since $\rho=V_{0}N=N/\rho_{0}$ (we define $\rho_{0}=1/V_{0}$, the density for a single cell within the tissue volume), $\Delta\propto 1-R^{\rho/\rho_{0}}$.

{\fp Thus, the functional form relating the CONDOR model's contraction parameter, $\Delta$, to seeding density, $\rho$, is (noting that the maximum possible $\Delta=1$),}
\begin{equation}
  \Delta = 1-R^{\rho/\rho_{0}}.
  \end{equation}


{\fp Without loss of generality, we define $R'=R^{\rho'_{0}/\rho_{0}}$, where $\rho'_{0}$ is a reference cell density. Since $R'$ and $\rho'_{0}$ are independent, we fix $R'=1/2$ and allow $\rho'_{0}$, to set the density scale. So,}
  \begin{equation}
 \Delta = 1-\left(\frac{1}{2}\right)^{\rho/\rho'_{0}}.
  \label{eqn:deltavsdens}
\end{equation}.

 With this choice, Eq. \ref{eqn:deltavsdens} has a linear form for $\rho\ll\rho'_{0}$, saturates for $\rho\gg\rho'_{0}$, and has the value $\Delta = 1/2$ when $\rho = \rho'_{0}$.

{\fp Eq. \ref{eqn:deltavsdens} may be inverted to determine $\rho$ from $\Delta$:}
\begin{equation}
  \rho = \rho'_{0}\ln(1-\Delta)/\ln(1/2).
  \label{eqn:densvsdelta}
  \end{equation}

\subsection{Computational method}

{\fp We determine cell orientations and positions using a simulated annealing approach to find the lowest energy configuration of the cell--spring network.} Bischofs and Schwarz have proposed that cells find an orientation
consistent with the minimum work needed to obtain that configuration
\cite{bischofs2003}, i.e. the ground state. With a sufficiently slow anneal, the global minimum energy is guaranteed \cite{kirkpatrick1983,szu1987}.

{\fp In the simulated annealing approach \cite{kirkpatrick1983,szu1987}, a set of Monte Carlo updates are made, where in each update we modify either cell positions or orientations (or both), for either single cells or groups of cells.} Cell positions and orientations are characterized using azimuthal and polar angles ($\theta$ and $\phi$) and their coordinates ($x,y$ and $z$). Updates are accepted or rejected according to a  Metropolis scheme, with probability $\exp(-\Delta E/\eta)$, where $\eta$ is slowly reduced according to either an exponential or fast schedule. $\Delta E$ is the change in energy due to an update.

{\fp During reorientation, both angles are updated either from a Lorenzian distribution \cite{szu1987} or a stepped distribution.} For the Lorenzian
distribution, angles are updated
according to $\theta \rightarrow \theta' = \theta + \Delta\theta
\tan\left(\pi(r-1/2)\right)$. Here $r$ is a random variate with range $[0,1)$, computed
  using the ran2 function from Press {\it et al.}
  \cite{press1992}. The alternative step function update is
$\theta \rightarrow \theta' = \theta + \Delta\theta (r-1/2)$, where
$\Delta\theta$ characterizes range. Additional updates with
$\Delta\theta' = \Delta\theta/10$ and $\Delta\theta'' = \Delta\theta/100$ are
also used. Similar updates are made for $\phi$.

{\fp A major difference between this and previous work
\cite{bischofs2003,bischofs2006} is that cells can move to any location, thus allowing structural changes of the tissue to be explored.} Lorenzian and step updates are made for $x,y$ and $z$. A fixed penalty is imposed on cells that move outside the working area of the mold. Within tethering regions, cells are not permitted to move, but can reorient.

{\fp  Annealing is carried out using several anneal speeds, two anneal
  schedules and fixed/variable update ranges, with the lowest energy configuration from the different attempts selected for analysis}. Various annealing schedules were used, including fast annealing and exponential annealing. The
  initial $\eta$ was selected by increasing $\eta$ with
  fixed update range until a high acceptance rate was
  achieved. Various schemes for selecting update range were used,
  including scaling of $\Delta\theta$, etc, with $\eta$ so that updates are
  accepted with roughly equal rate throughout the anneal. For variable
  update range, the algorithm is considered to be finished when the
  ranges are less than a threshold, whereas for fixed range the
  algorithm finishes once the acceptance rate drops below a preselected value. Typical anneal times are a few CPU days.
  
  {\fp Following the anneal, the lowest energy configuration is used to calculate tensions in the springs, and contraction ratios.} For consistency between contractile networks with different length scales, we introduce a dimensionless tension, $\bar{T}=\sum T/\kappa_{\rm NN} l_0$. This measurement describes the tension in the environment of a cell.

\subsection{Experimental data}

We also carry out the task of validating our microscopic biophysical model, by comparing with experimental data that have been published elsewhere \cite{orourke2015,orourke2017}. We have not performed any new experiments for this paper, and emphasize that the primary purpose of this paper is to introduce a microscopic model of cell-ECM interactions in tissues. A summary of the raw experimental data from Refs. \cite{orourke2015,orourke2017} that we have used in our comparisons can be found in the supplementary material. 

{\fp Full experimental details have been reported previously, however for convenience we provide a brief summary here \cite{orourke2015,orourke2017}.} Cells (C6 rat glioma cells or primary rat glial cells) were mixed with a neutralized solution containing acid-solubilized 2mg/ml Type I rat tail collagen and minimum essential medium then cast within multiwell plates (for contraction assays) or bespoke tethering molds (for alignment experiments). Cellular collagen gels were allowed to set for 10 minutes then immersed in culture medium and incubated for 24 hours at 37$^o$C. During incubation, time natural cell-matrix interactions resulted in gel contraction and, in tethered gels, tension-mediated alignment of cells and matrix. Cellular gels were fixed and analyzed using immunofluorescence staining and confocal microscopy to visualize cell shape and position, with quantification of cellular alignment assessed using Volocity software (Perkin Elmer, Waltham,
MA, USA).    

{\fp There are two minor additional analyses carried out here.} We convert the area contractions in  Ref. \cite{orourke2015} to a linear contraction for comparison with our model, by assuming that the resulting gels are circular and obtaining the radius of that circle, with error bars updated accordingly. For the tissue reported in Ref. \cite{orourke2017}, we reanalyze experimental cell orientations to provide information on cell orientations across the whole sample for comparison with theoretical predictions. This is done from the original confocal micrograph using Volocity.

\section{Results}
\label{sec:results}

\subsection{Untethered tissues in wells}

\begin{figure}
    \includegraphics[width=85mm]{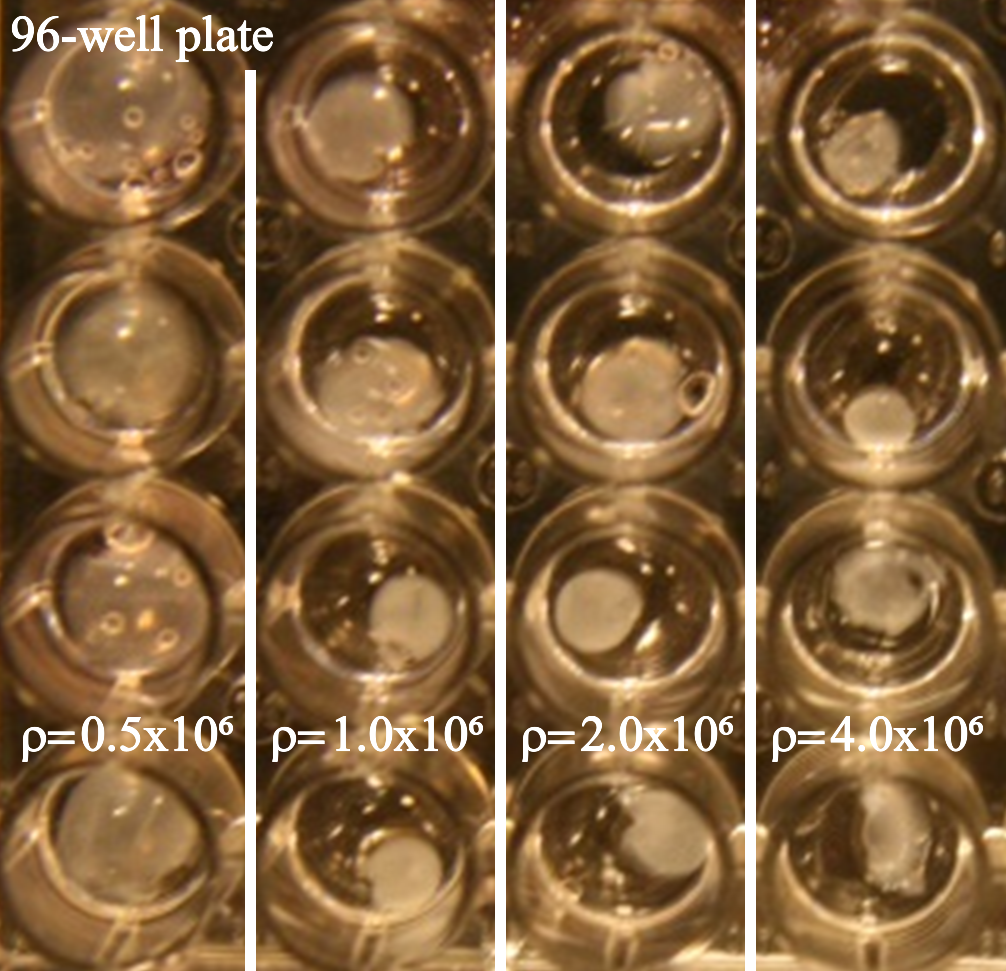}
    \caption{Image of free-floating artificial neural tissues grown in circular wells (modified from Ref. \cite{orourke2015}), for visual comparison with results from our theoretical model in Fig. \ref{fig:cellorientationwell}. Contraction increases with density (left to right).}
  \label{fig:experiment96well}
  \end{figure}

{\fp In this section, we apply our microscopic model to one of the simplest 3D artificial tissue constructs, which has no tethers and can be said to be free-floating.} Free-floating tissue can be grown in circular wells (96-well plates). For convenient comparison with results from our microscopic model Fig. \ref{fig:experiment96well} displays an image reproduced from Ref. \cite{orourke2015}, showing how the shape of the tissue is affected by the presence of cells. Initially, the whole well is filled with a mixture of collagen and cells. During the incubation period, the cells cause the tissue to contract. Cell-induced contractions in untethered artificial tissues increase with cell density. 

\begin{figure}
    \includegraphics[width=85mm]{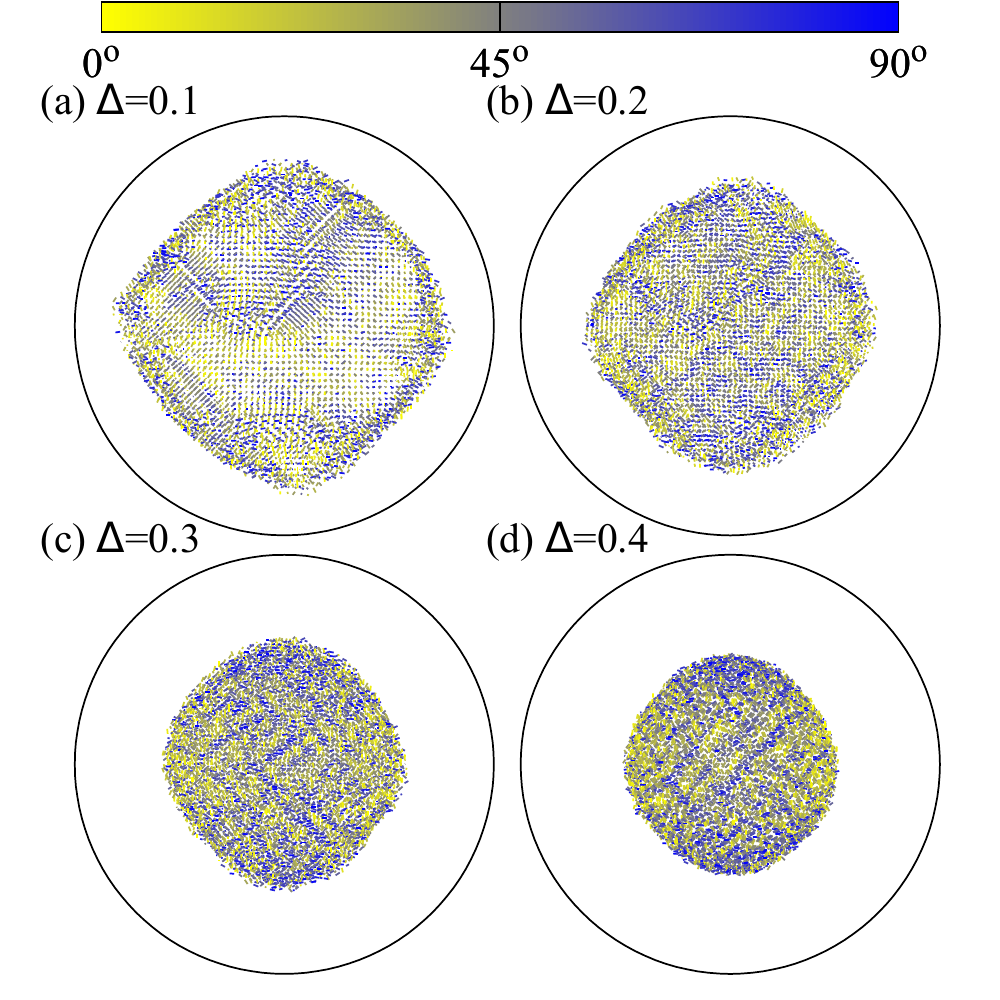}
    \caption{Simulations using our microscopic model show that free-floating artificial glial tissue contracts with increased $\Delta$, with cell orientations showing domain boundaries. Data correspond to 96-well plates. Black solid lines denote the perimeter of the well. At the start of the simulation, cells are uniformly distributed within the well. The tissue within the molds is three-dimensional, but shown in profile.}
    \label{fig:cellorientationwell}
    \end{figure}

{\fp Our computational simulations of self-organization in free-floating (i.e. untethered) artificial tissue in cylindrical wells, show increased contraction as $\Delta$ gets bigger.} Figure \ref{fig:cellorientationwell} shows top view theoretical calculations for the cell orientation in free-floating artificial glial tissue grown in 96-well plates for a variety of $\Delta$ values. For the 96 well plates, the ratio of initial gel height to diameter is $\sim 0.1$. Domains are visible in the cell orientations. Shapes are similar to the experiments shown in Fig. \ref{fig:experiment96well}.
  
\begin{figure}
    \includegraphics[width=85mm]{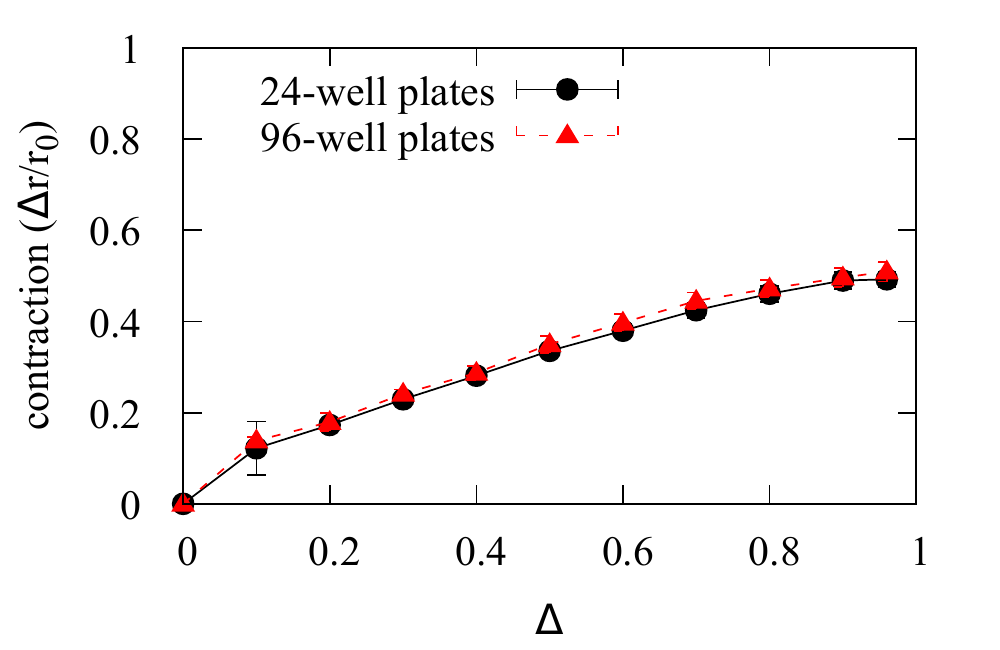}
    \caption{Simulations using our microscopic model predict that contraction increases with $\Delta$ for free-floating (i.e. untethered) artificial glial tissue, with contraction saturating at around 50\%. Calculations correspond to 96-well plates.}
    \label{fig:contractvsdelta}
    \end{figure}

{\fp From these calculations, the contraction, $\Delta r/r_{0}$ vs. $\Delta$ can be determined for free-floating gels in cylindrical wells, and is found to increase monotonically, saturating at $\sim 50\%$ contraction.} Here $\Delta r$ is the change in radius, and $r_{0}$ the initial radius. Figure \ref{fig:contractvsdelta} shows $\Delta r/r_{0}$ for simulations with various $\Delta$ values for free-floating (i.e. untethered) artificial glial tissue grown in circular 24-well and 96-well plates \footnote{For those not familiar with cell culture terminology these are standardized trays with circular depressions for growing and analyzing cells and tissues, with the number referring to the number of depressions within the tray.}. The ratio of initial thickness to well diameter is $\sim 0.2$ in the 24-well plates. Contractions in the horizontal plane are shown, since they correspond to measured experimental contractions from Ref. \cite{orourke2015}. For free-floating gels, contraction tends towards $\sim$50\% as $\Delta\rightarrow 1$. The depth of gel does not contribute strongly to the size of the contraction. 
  
{\fp   A linear interpolation of simulation results in Fig. \ref{fig:contractvsdelta}, combined with experimentally measured contractions from Ref. \cite{orourke2015}, are used to determine how $\Delta$ relates to $\rho$.} Experimental data from Ref. \cite{orourke2015} exist for two cell types, the C6 glioma line and primary rat astrocytes.  Ranges of $\Delta$ are determined from the experimental variance. Calculations corresponding to 96-well plates were used.

\begin{figure}
    \includegraphics[width=85mm]{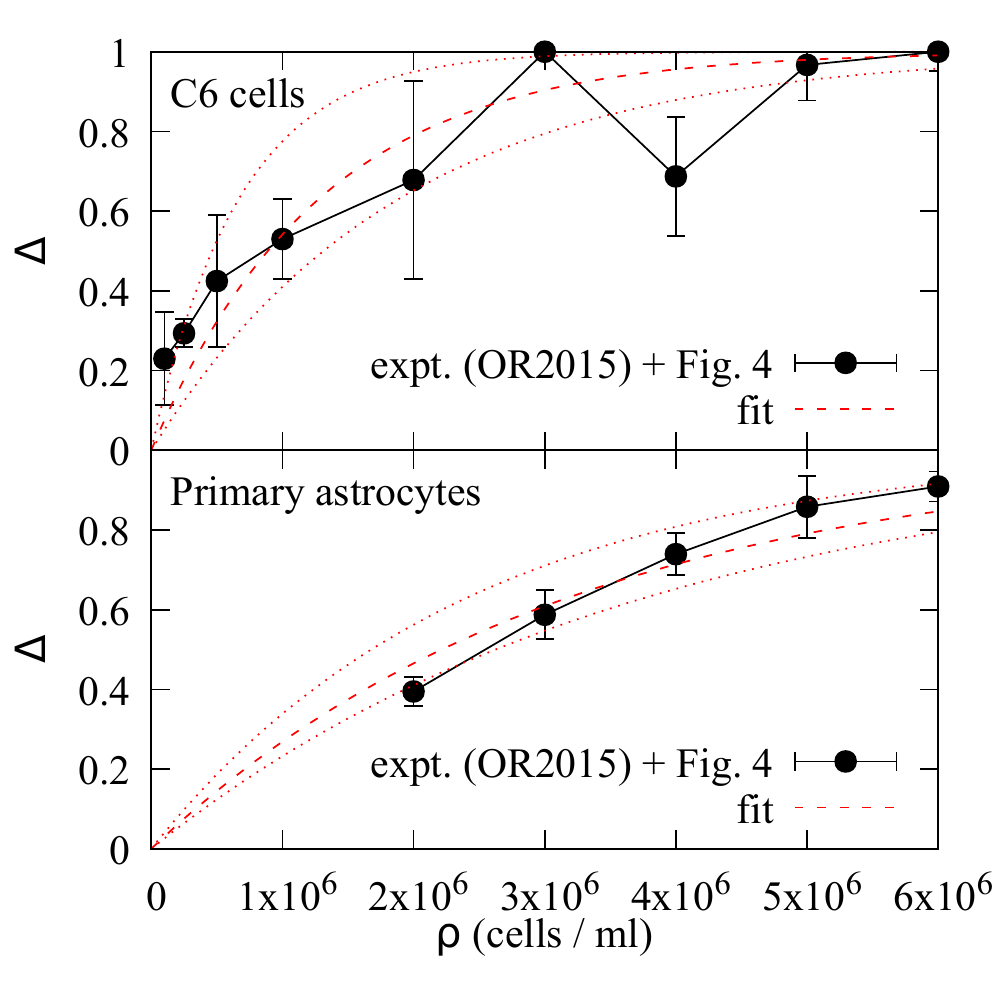}
    \caption{We make a linear interpolation of data in Fig. \ref{fig:contractvsdelta} combined with experimental data reported in Ref. \cite{orourke2015} (labeled OR2015) to determine how $\Delta$ relates to $\rho$, with ranges of $\Delta$ determined from the contraction in 96-well plates, and the resulting plot is used to determine $\rho'_{0}$ from $\Delta$ using a least squares fit of Eq. \ref{eqn:deltavsdens}. The dashed line shows the fit and the dotted line the 95\% confidence interval on the fit.}
    \label{fig:deltavsrho}
    \end{figure}

{\fp The resulting data relating $\Delta$ to $\rho$ are used to determine $\rho'_{0}$ using a least squares fit, showing that the functional form of Eq. \ref{eqn:deltavsdens} (derived by us) is consistent with the experimental data from Ref. \cite{orourke2015} (Fig. \ref{fig:deltavsrho})}. For C6 cells, $\rho'_{0}=0.88(9)\times 10^{6}$cells/ml with a lower bound on the 95\% confidence interval of $\rho'_{0,-}=0.46(4)\times 10^{6}$cells/ml and upper bound $\rho'_{0,+}=1.31(3)\times 10^{6}$cells/ml. For primary astrocytes, $\rho'_{0}=2.15(1)\times 10^{6}$cells/ml with confidence interval $\rho'_{0,-}=1.67(7) \times 10^{6}$cells/ml to $\rho'_{0,+}=2.62(4)\times 10^{6}$cells/ml. We note that data for C6 cells have larger error bars and the fit is less good, with large uncertainties on the values of $\rho_{0}'$.
  
\begin{figure}
    \includegraphics[width=85mm]{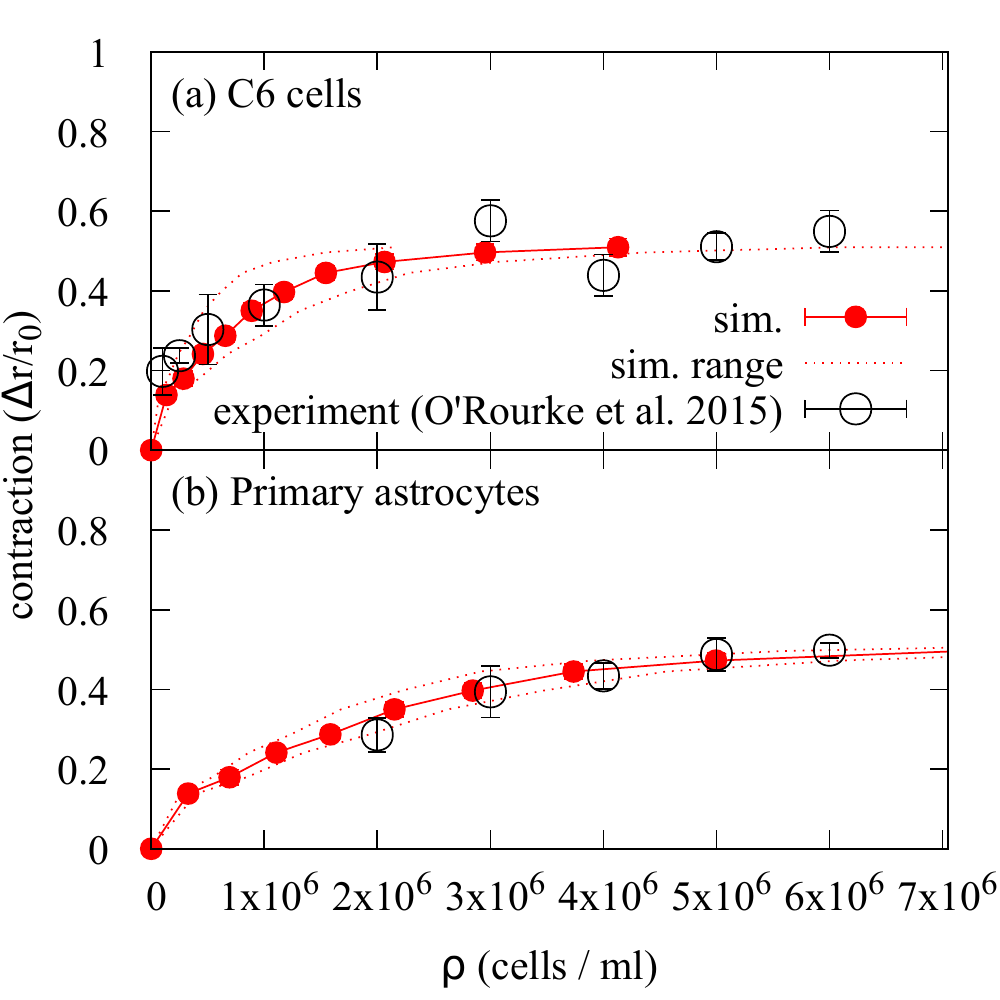}
    \caption{Simulations of free-floating artificial tissue in wells (reported here) match experimental contraction data (reported in Ref. \cite{orourke2015}, Figs. 3 and 5).}
    \label{fig:contractionvsdensity}
    \end{figure}

  {\fp Using these values of $\rho_{0}'$ to set the density scale, it can be seen that our microscopic model predicts contraction vs. density curves that agree well with experiment (Fig. \ref{fig:contractionvsdensity}).} Of course, this is to be expected, since a combination of theoretical and experimental values were used to find $\rho'_{0}$, however agreement would not be possible if the the functional form in Eq. \ref{eqn:deltavsdens} were not reasonable. Our theoretical predictions and the experimental measurements from Ref. \cite{orourke2015} are both shown. We note that the area covered by the tissue is measured in Ref. \cite{orourke2015}, and we have converted this to an effective radius for comparison with our results. 
  
\subsection{I-shaped tethers}

\begin{figure}
    \includegraphics[width=45mm]{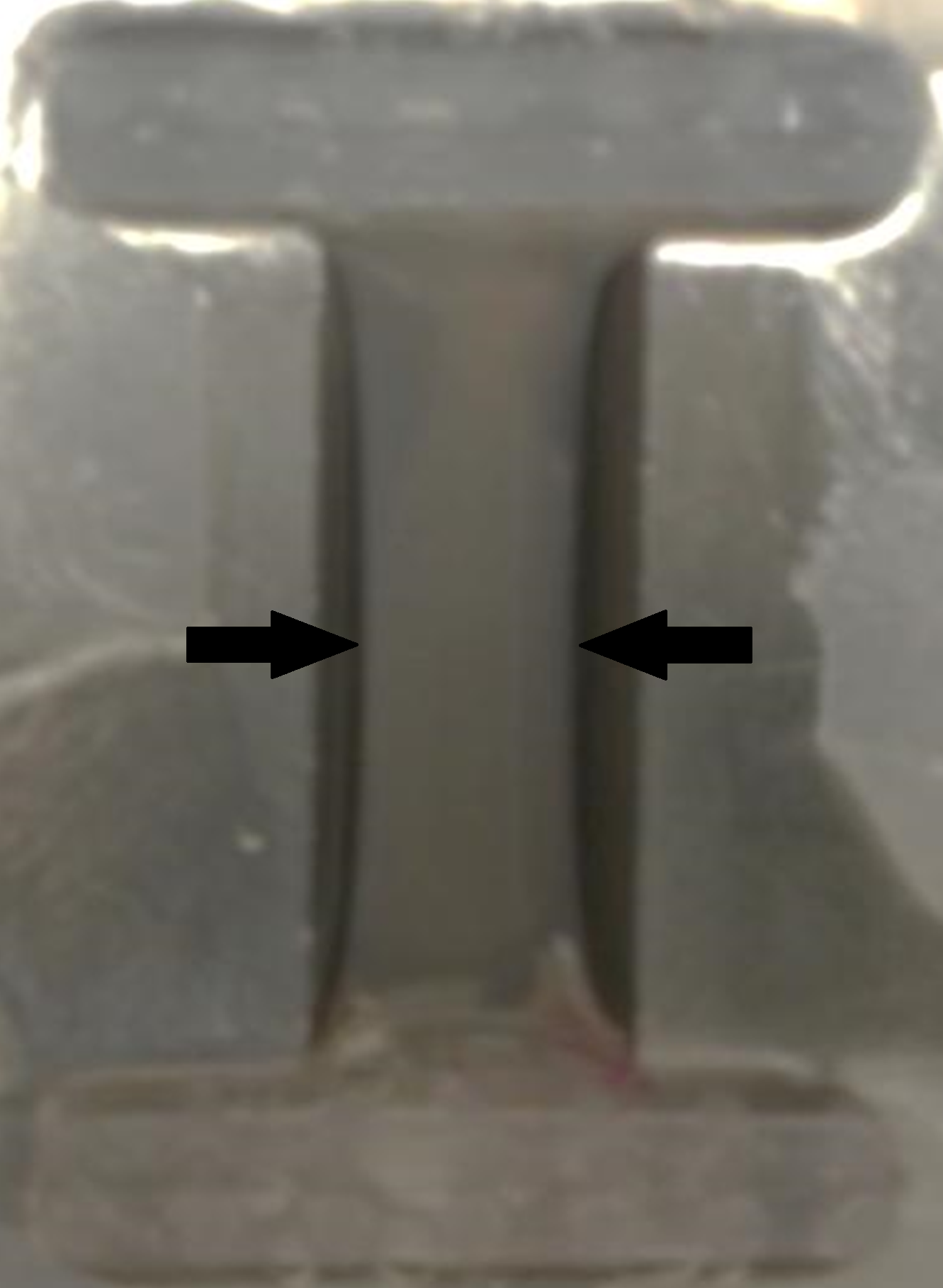}
    \caption{Image of tissues grown in an I-shaped mold, where reshaping of the ECM is modified by tethers (reproduced from Ref. \cite{orourke2015}) for visual comparison with simulations from the theoretical model in Fig. \ref{fig:soishape}. Cells are self-organized at the center of the sample (not shown). The contracted width of the tissue (highlighted with arrows) is reported in Ref. \cite{orourke2015}.}
  \label{fig:experimentishape}
  \end{figure}

 {\fp In this section, we apply our biophysical model to the self-organization of engineered tissue in I-shaped clamps.} Tethers can be used to guide tension in the engineered tissue. For convenient comparison with our simulations, Fig. \ref{fig:experimentishape} shows an image of tissue grown in an I-shaped mold reproduced from Ref. \cite{orourke2015}. The mold has clamping bars at the top and bottom of the image, within which there are multiple tethering points. In contrast to, for example, MEMS devices \cite{legand2009}, the tethers that are simulated in this paper are passive, and do not exert any tension additional to that generated by cells. The biggest contraction of the tissue is towards the center of the sample, and there is also self-organization of cells in this region (not shown).  The tether restricts contraction towards the top and bottom of the sample.

\begin{figure}
     \includegraphics[width=85mm]{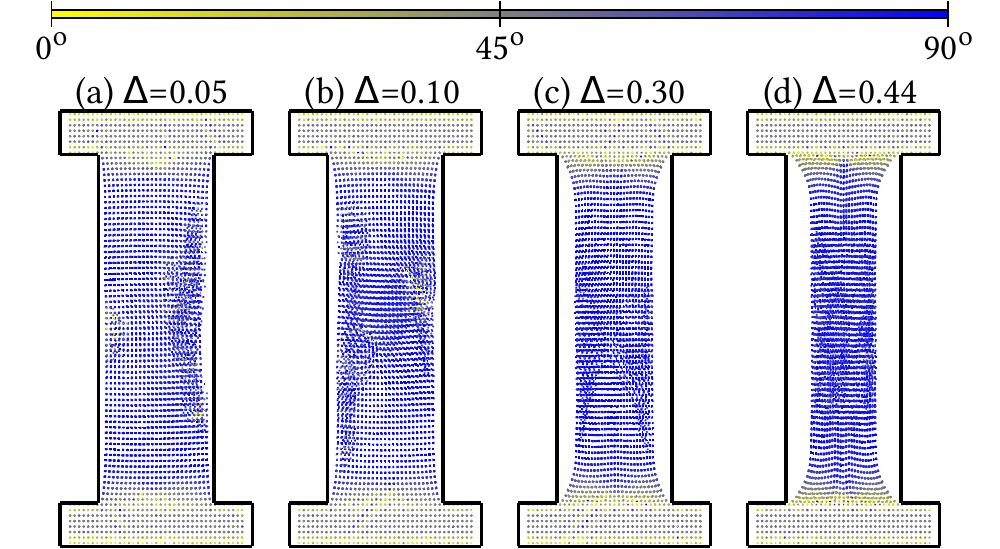}
     \caption{Our theoretical predictions for the shape of artificial glial tissues tethered in an I-shaped clamp match the shapes of experimental tissues from Ref. \cite{orourke2015} and reproduced in Fig. \ref{fig:experiment96well}. Cell alignments are also shown. All panels correspond to C6 cells with $\rho = 4\times 10^{6}$ cells/ml. (a) $\Delta=0.05$ to 1$\mu$g/ml anti-CD49; (b) $\Delta=0.1$ to 0.3$\mu$g/ml anti-CD49; (c) $\Delta=0.3$ to 0.1$\mu$g/ml anti-CD49; (d) $\Delta=0.44$ , without anti-CD49.}
     \label{fig:soishape}
    \end{figure}

{\fp Tissue shapes predicted by our microscopic model agree well with experimental results from Ref. \cite{orourke2015}}. Figure \ref{fig:soishape} shows our theoretical predictions (shape and cell orientations) for artificial glial cells in an I-shaped clamp. These can be compared with experimental results (Fig. \ref{fig:experimentishape}) from Ref. \cite{orourke2015} (shape only).  C6 cells of density $\rho = 4 \times 10^{6}$cells/ml form the artificial tissue.

\begin{figure}
    \centering
    \includegraphics[width=85mm]{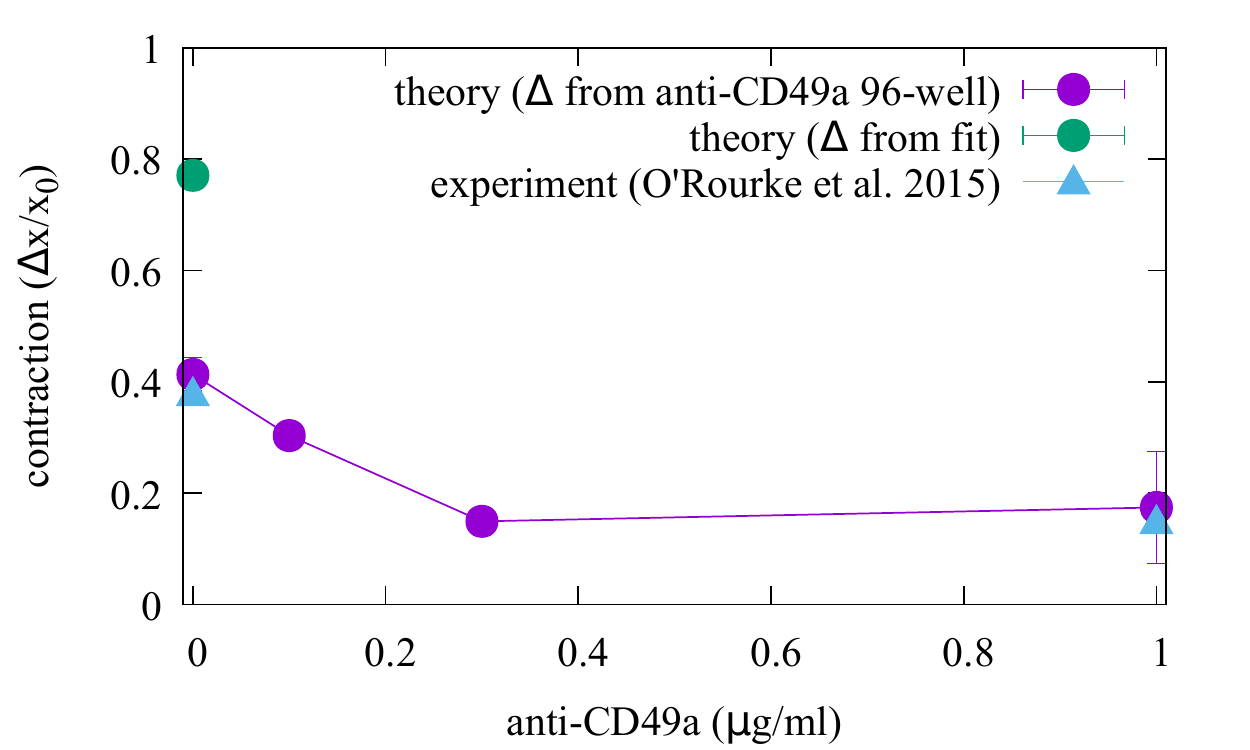}
    \caption{Contractions of tissue in I-shape tethers from our theoretical model (reported here) show good agreement with experimental data (reported in Ref. \cite{orourke2015}), if 96-well data are used to set $\Delta$. Points marked `experiment' correspond to contraction of tissue tethered in an I-shaped clamp for hydrogels, made from C6 cells and infused with various concentrations of anti-CD49 antibodies. Agreement is poor if Eq. \ref{eqn:deltavsdens} is used to set $\Delta$.}
    \label{fig:ishapecontractionvsexpt}
\end{figure}

{\fp Agreement is good between our theoretical predictions of contraction and experimental measurements from Ref. \cite{orourke2015} (Fig. \ref{fig:ishapecontractionvsexpt}).} Experimental data correspond to C6 cells tethered in an I-shaped clamp and infused with various concentrations of anti-CD49 antibodies. Anti-CD49 antibodies are used to block integrins, and therefore the ability for cells to bind to the ECM, leading to a significant decrease in the contraction of the tissue. If 96-well data (reported in Fig. 1 of Ref. \cite{orourke2015}) are used in conjunction with Fig. \ref{fig:contractvsdelta} of this work to determine $\Delta$ then agreement between simulation and experiment with respect to contraction is excellent. We note that contraction of the hydrogel in the I-shaped clamp is smaller than predicted if Eq. \ref{eqn:deltavsdens} is used to set $\Delta$. The consequences of this for calibration of the CONDOR model will be discussed later. 

\begin{figure}
     \includegraphics[width=85mm]{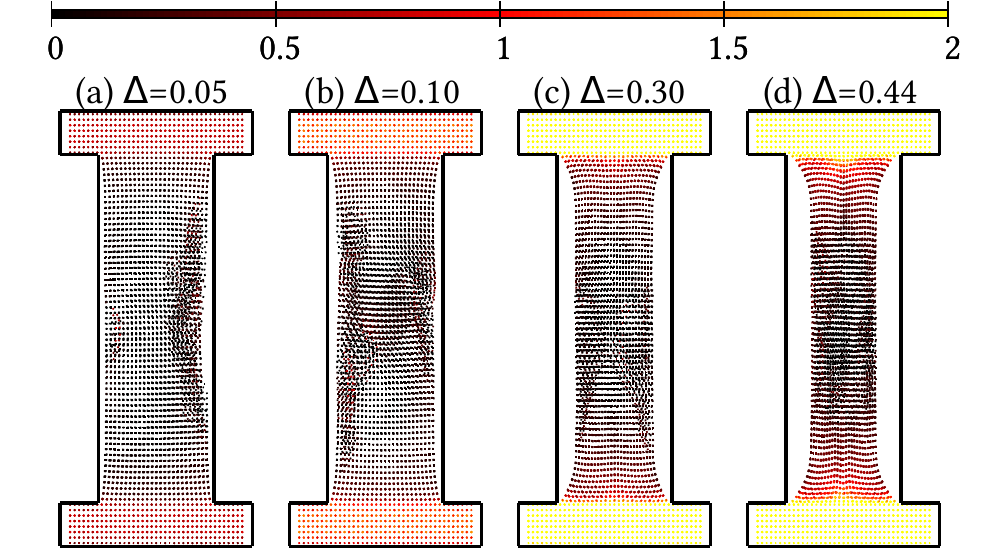}
     \caption{Our simulations of tissue in an I-shaped clamp show regions of high tension around the clamps. The color represents $\bar{T}$, with yellow representing regions of highest positive tension,
red regions of moderate tension and black regions of lowest
tension.}
     \label{fig:tensionishape}
    \end{figure}

{\fp An important benefit of the microscopic model presented here is the ability to gain insight
into quantities that would be difficult or impossible to determine
experimentally, such as tension.} In
Fig. \ref{fig:tensionishape}, we show theoretical predictions of the
tension in free-floating artificial glial tissue grown in an I-shaped mold. In well-aligned regions, tension is low. Frustration associated with rapid changes of alignment with position leads to high tension close to the clamps. This high-tension region is Delta-shaped. Insight into difficult-to-measure quantities could be used when designing bespoke clamps for artificial tissue with specific structures. 
  
\subsection{Pillar tethers}
  
 {\fp In this section, we discuss the application of our microscopic biophysical model to self-organization in molds where pillars tether the tissue.} The relevant mold is a rectangular well with four pillars at each of the corners of the mold. As the hydrogel sets, the pillars hold the corners of the tissue in place, and direct tension as the tissue contracts under the influence of cells. Again, we note that the pillars are fixed and do not actively exert tension.
  
\begin{figure}
  \begin{center}
    \includegraphics[width=0.45\textwidth]{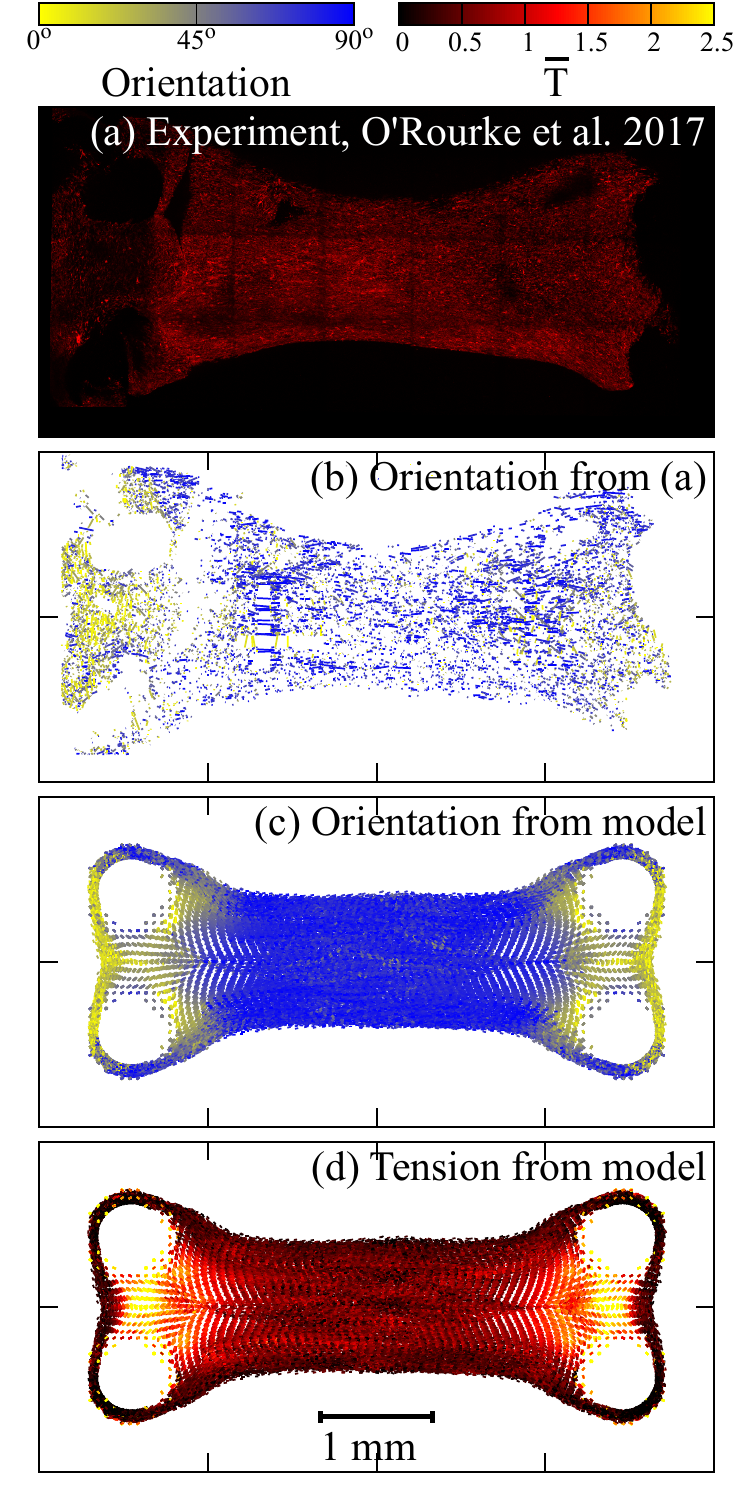}
    
  \end{center}
\caption{Our simulations agree well with both shape and cell orientation of artificial tissue tethered with vertical pillars reported in Ref. \cite{orourke2017}. (a) A confocal micrograph of the engineered tissue reproduced from Ref. \cite{orourke2017}. Tethering holes are visible on the left of the image and there is a characteristic bow-tie shape; (b) Cell orientations extracted from the micrograph using Volocity; (c) Theoretical results show self-organized cell alignment along the long axis and also along the short axis between tethers, with a region of various orientations at the interface between the two, as in the experimental data; (d) Moderate tension is found in the central region, which is aligned along the long axis. Regions of low tension are seen at the edge of the
  sample. Mixed tensions are found in the triangular region of mixed
  orientations. Regions of highest
  tension correspond to the region aligned along the short axis between the
  pillars.}
\label{fig:pillararrangement}
\label{fig:experimental}
\end{figure}

{\fp In a mold where pillars are used to tether the hydrogel, interactions between cells and ECM also lead to narrowing of the artificial tissue, as seen in confocal micrograph images (Fig. \ref{fig:experimental}(a), which is reproduced from Ref. \cite{orourke2017}).} In panel (a), cells are stained red and the background with no cells is black. The artificial tissue starts as an untensioned rectangular type I collagen gel seeded with glial cells tethered by a set of four pillars, two to the left and two the the right of the image. These tethers show up as holes in the tissue sample after it is removed from the mold. It obtains a bow-tie shape after 24 hours of incubation. C6 cells of density $\rho = 4 \times 10^{6}$cells/ml form the tissue.

{\fp In order to determine the predictive power of our simulations, we determine cell orientations by analyzing the full confocal micrograph (Fig. \ref{fig:experimental}(b)).} The analysis highlights that the pillar-tethered hydrogel has a triangular (Delta) region just to the right of the two left-hand pillars, a region where cells are aligned along the short axis between the pillars and a region of glial cells
aligned along the long axis towards the center of the sample. Volocity (PerkinElmer, Waltham,
MA, USA) is used to determine cell alignments from the image in Panel (a). Yellow indicates cells aligned along the
short axis, and blue indicates cell alignment along the long axis. Approximately 10\,000 cells are
identified in the confocal micrograph image. We note that the sample has been compressed under a cover slip. 

{\fp Our computer simulations of cells in a pillar-like mold (Fig. \ref{fig:pillararrangement}(c)) reproduce key features seen in the analyzed confocal image (Fig. \ref{fig:pillararrangement}(b)): regions of self-organized cell
alignment can be seen between the pillars and along the long axis at the center of the sample, and regions of mixed
orientations are found where regions of different alignment meet.} The positions of cells in the CONDOR model are
fixed around the pillar, but permitted to change angle. All other cells are allowed to relax their positions. Results are
shown for $\Delta = 0.8$ (which is consistent with the  density of $4\times 10^{6}$cells/ml in the experimental data) and $l_{0} = 100\mu\mathrm{m}$ (panel
(c)). Calculations contain 59\,891 cells. 

{\fp Regions of highest dimensionless tension, $\bar{T}$ are associated with regions of rapidly changing cell orientation, as in the tissues simulated in I-shaped molds.} The strongest tension can be found in the small region between
the pillars, extending into a roughly triangular shape of higher tension along the long
axis (Fig. \ref{fig:pillararrangement}(d)). So again, higher tension is associated with regions of rapidly changing alignment. The color represents the dimensionless value $\bar{T}$, with yellow representing regions of highest positive tension, red regions of moderate tension and black regions of lowest tension. Contraction at the center of the sample is consistent between experiment and theory.

\section{Discussion and conclusions}
\label{sec:discussion}

{\fp In this paper, we have introduced the CONDOR model, a microscopic biophysical model that combines contractile networks and force-dipoles, to describe self-organization due to the feedback between cells and the extracellular matrix in tissues.} In the CONDOR model, a network of springs represents ECM and the active forces generated by cells are represented by an additional contraction with a dipole symmetry. The lowest energy configuration of the CONDOR model is found using simulated annealing. Results from our CONDOR model match experimental measurements of artificial tissue from Refs. \cite{orourke2015} and \cite{orourke2017} both on a microscopic level from cell alignments, and on a macroscopic level for the shapes of tethered tissues. We consider the ability to make macroscopic scale predictions from a microscopic scale model to be an important step forward for modeling artificial tissue.

\subsection{Model validation and calibration}

{\fp We have validated our microscopic biophysical model against existing experimental data that can be found in Refs. \cite{orourke2015} and \cite{orourke2017}.} Our model performs well for a variety of situations, including free-floating gels in 96-well plates, gels clamped in I-shaped molds, and gels tethered using pillars. It is capable of good agreement with both the shape of the tissue and the orientations of cells within the tissue.

{\fp Calibration of the microscopic model requires determination of $\rho_{0}'$, and is more challenging.} One strategy is to fit the outputs of the model to data from free-floating gels, and we have done this using the data reported in Ref. \cite{orourke2015}. Determination of $\Delta$ from multi-well plates is particularly sensitive to statistical fluctuations in experimental results, due to the slow change in contraction with $\rho$, especially at larger $\rho$ values. This indicates that very high accuracy experimental data are required to determine $\rho'_{0}$ and thus calibrate the model. 
The experiments in \cite{orourke2015,orourke2017} were carried out with the usual care to control confounding variables such as passage (which is integer and can be matched exactly) and confluency, which is controlled to around 60-70\% \footnote{For those not familiar with cell culture terminology, confluency refers to the cell density in the flask used to grow cells. At a low confluency the cells can multiply freely. At high confluency, crowding of cells slows their multiplication rate. Confluency also affects contraction. Ideally cells should be used at 60-70\% confluence. Eventually, cells need to detached from the flask and either used for experiments or divided between a number of flasks to expand and increase in number. Each time cells are divided between flasks, passage number is incremented. For many cell types, growth rate decreases with increased passage number and again contraction can also be affected.}.  
However, even with this care, there can be large variations in tissue contraction, due, not only to these variables, but also since the tissue can sometimes adhere to the sides of the mold in addition to the tethering points. This variation leads to sizable uncertainties in the value of $\rho'_{0}$ for a particular cell type. 

{\fp An alternative strategy for calibrating the microscopic model is to determine $\Delta$ for specific batches using the contractions seen in multi-well plates.} This strategy leads to very close agreement between theory and experiment if tethered molds are seeded with precisely the same cell batch.

\subsection{Context}

{\fp Our microscopic biophysical model combines contractile networks and force dipoles, to predict self-organization and reshaping of tissue due to the microscopic active forces between cells and the ECM.} We are unaware of other microscopic biophysical frameworks for modeling this process. However, we note similar approaches.

{\fp Pure force-dipole models do not incorporate the feedback between self-organization and reshaping driven by active forces that are found in our microscopic model (and real tissues).} The symmetries of the active forces in our model are similar to force dipoles. Force dipoles have been a popular approach for describing the interactions between cells and the ECM \cite{schwarz2013,bischofs2003,bischofs2004,bischofs2005,zemel2006,schwarz2002,bischofs2006,zemel2007,friedrich2011}. The force-dipole approach is sophisticated and has provided insight into the reasons that cells orient, especially in response to external stresses. Repulsive terms leading to finite compressibility of the viscoelastic ECM are not present. To stop collapse, cells in these theoretical models are constrained either to fixed positions, or small movements about such fixed positions, so reshaping of the tissue cannot be described.

{\fp With the exception of the microscopic model presented here, we are not aware of any contractile network models that have been modified to include the active forces between cells and ECM.} In our microscopic model, the ECM is represented by a contractile network model. Contractile network models are common in biophysics, and can be used to represent many elastic media \cite{boal}. For example, contractile network models have be used to describe the shapes of both tethered single cells and bulk tissues \cite{bischofs2008,guthardttorres2012,bischofs2009}. 

{\fp In the context of our microscopic model, continuum mathematical models provide a complementary, yet distinct, approach for describing self-organization in tissues.} We note continuum approaches to the alignment of stress fibers in artificial tissues, which can accurately predict the shapes of artificial tissue grown in MEMS systems \cite{deshpande2006,pathak2008,legand2009}. These approaches are concerned with actin fibers and assume cell alignment along such fibers. We are not aware of any other continuum approaches that can describe the feedback between reshaping and self-organization in tissues.

{\fp Our CONDOR model has several key strengths, including:}
\begin{enumerate}
\item Accurate description of the self-organization and reshaping of tissue that is driven by the feedback between cells and ECM.

\item Predictive capability for the shapes of tissues and the orientations of cells in those tissues.

\item Simplicity allows for simulation of large numbers of cells allowing macroscopic tissue sizes to be approached.

\item Generalizability to different symmetry arrangements by adding higher order multipolar corrections or other alternative symmetries to $f(\theta)$ to allow the investigation of other biophysics.

\item Extensibility to disorder, which can be achieved by including random bonds or
by making corrections to the equilibrium distances of bonds. 

\item  Similarities to models common in statistical phsyics (such as the Heisenberg model of spin alignments, and models of liquid crystals), allowing the extensive machinery of statistical physics to be used to solve the microscopic model (e.g. Monte Carlo methods).
\end{enumerate}

\subsection{Potential applications of the model}

{\fp Specific patterns of cellular self-organization are important in many tissues, making it challenging to grow artificial tissues that properly represent those in living organisms. } The use of artificial or engineered tissue for regenerative medicine has the potential to significantly improve human health \cite{obrien2011} and provide new biological models for drug testing \cite{orourke2017}. The cells in real tissues are typically arranged in ways that may be difficult to reproduce artificially. For example, this is a particular problem for nerve tissue, which must be well-ordered to function properly. In the nervous system, neurons are typically guided by glial cells, so well-ordered glial cells in artificial tissue could lead to well-directed neuron growth for, e.g., nerve repair. For such applications, favorable characteristics are high levels of cell alignment and cross-section that is uniform along the length of the  tissue, while regions of misalignment around tethering points should be minimized.

{\fp  Design of artificial tissue with realistic characteristics is something that is very difficult and time consuming to do experimentally using trial and error, so the ability to predict outcomes would be extremely valuable.} 
The CONDOR model described here is fast enough to  assist the rational design of experimental molds and scaffolds that guide artificial tissue growth. The comparison with experiment carried out here indicates that for a design process, a broad range of $\Delta$ should be used to obtain indicative results for specific mold designs.

\subsection{Future prospects}

{\fp  Future work will increase sophistication of the CONDOR model to improve predictions of self-organization in tissue.} There are several ways in which the model could be extended:

1. Improved cell response to tension in the ECM: The cells in the CONDOR model are always elongated (polarized), however experimental systems have regions where cells are not elongated, particularly within the Delta region, so the contraction of the sample may be over-estimated. The CONDOR model could be modified so cells are unpolarized in regions of low tension.

2. Increased sophistication to improve the description of the ECM and analyse the role of disorder: For example, the polymers making up the ECM have many signatures of disorder.  Randomized spring networks could therefore improve the description of collagen networks in the ECM. An intriguing possibility for further physics investigation, is the existence of phase transitions between self-organization and disorder in the CONDOR model. 

3. Inclusion of time dependence: Cell representations in the model could be extended to include dynamic responses. This would allow the study of time-dependent ECM remodeling, the evolution of samples during their growth phase \cite{orourke2015}, or oscillatory strain stimuli (e.g. \cite{de2007,de2008,safran2009}).
 
Our microscopic model is amenable to all of these extensions and work is currently in progress. Overall, we expect that a wide range of interesting biophysics can be studied within extensions to the CONDOR model introduced here.

\section*{Acknowledgments} The authors would like to thank Dr. Calum MacCormick for useful discussions. The authors have no competing interests.

\bibliography{references}
\bibliographystyle{unsrt}

\end{document}